\documentclass[11pt]{article}

\usepackage[dvips]{graphicx}
\usepackage{amsmath,amssymb}

  \newcommand{\be}{\begin{equation}}
\newcommand{\ee}{\end{equation}}

\title{\large{\textbf{FLUCTUATION PROPAGATOR AND HEAVY QUARK
DIFFUSION}}}
\begin{document}

\date{ }
\maketitle

\begin{center}
\vspace*{-1cm}
\author{ \small{B. Kerbikov}}

\small{\itshape{State Research Center\\
Institute of Theoretical and Experimental Physics,\\
Moscow, Russia}}
\end{center}

\begin{center}

\footnotesize{The quark fluctuation propagator is evaluated.\\
It defines
the diffusion coefficient in the vicinity of the phase transition\\
and the gradient term in the Ginzburg-Landau functional.

\vspace{0.5cm} Contribution to the proceedings of the Rencontres de
Moriond on QCD and Hadronic\\ Interactions, La Thuile, Italy, March
2007.}

\end{center}

\section*{\normalsize{1. Introduction}}
Recent experiments performed at RHIC have revealed an unexpectedly
large heavy flavor suppression. This result may indicate that the
heavy quark diffusion coefficient is anomalously small in the
vicinity of the critical line of the QCD phase diagram in the
$(\mu,T)$ plane. The diffusion coefficient enters also into the
Ginzburg-Landau functional via the term proportional to the
$<A_{\mu}^{2}>$ condensate [1]. Comparison with the lattice
calculations of $<A_{\mu}^{2}>$ shows that due to some dynamical
reasons the diffusion coefficient is much smaller than the value
given by the simple Drude formula [1]. It is well known that the
diffusion coefficient may become small in the fluctuation regime and
turns zero at the Anderson localization edge.

The transport coefficients are expressed in terms of the
time-dependent propagator. Below we draw the fluctuation quark
propagator (FQP) in its simplest form as a preliminary step in the
investigation of the heavy quark dynamics in the vicinity of the
phase transition. Our derivation closely follows the guidelines of
the condensed matter physics [2].

Denoting the FQP as $L(\vec{p},\omega)$ we may define it as

$$L^{-1}(\vec{p},\omega)=-\frac{1}{g}+F(\vec{p},\omega),\eqno (1)$$

$$F(\vec{p},\omega)=\sum_k G(\vec{k},k_4) G(\vec{p}-\vec{k},\omega-k_4), \eqno (2)$$

Here g is the coupling constant with the dimension $m^{-2}$, the sum
over k implies the momentum integration, Matsubara and discrete
indices summation, $G(\vec{k},k_4)$ in the thermal Green's function
which reads

$$G(\vec{k},k_4)=\frac{i}{\vec{\gamma}\vec{k}+\gamma_4k_4-im+i\mu\gamma_4}, \eqno (3)$$
with $k_4=-\pi(2n+1)T, T=\beta^{-1}$. We shall compute
$F(\vec{p},\omega)$ in the approximation of long-wave fluctuations

$$F(\vec{p},\omega)\simeq A(\omega)+B(\omega) \vec{p}^{2}, \eqno (4)$$
First we compute the term $A(\omega)$. We have

$$tr\{G(\vec{k},k_4)G(-\vec{k},\omega-k_4)\}=2\left\{\frac{1}{\widetilde{k_4}^{2}+(\varepsilon-\widetilde{\mu})^{2}}+\frac{1}{\widetilde{k_4}^{2}+(\varepsilon+\widetilde{\mu})^{2}}\right\}, \eqno (5) $$
where tr is over the Lorentz indices,
$\varepsilon^{2}=\vec{k}^{2}+m^{2}$, $\widetilde{k_4}=k_4-\omega/2$,
$\widetilde{\mu}=\mu-i\omega/2$. The second term in (5) corresponds
to antiquarks. We shall omit it here only for brevity though as
shown in [1] the interplay of the quark and antiquark modes may
result in instability in the chiral limit. For the same reason we
put $m=0$. Performing the momentum integration around the Fermi
surface we obtain
$$A(\omega)=\nu\sum_{n\geq0}\frac{1}{\left(n+\frac{1}{2}+\frac{\omega}{4\pi T}\right)},\eqno (6)$$
where $\nu=\frac{2\mu^{2}}{\pi^{2}}$ is the density of states at the
Fermi surface for two quark flavors. To evaluate $B(\omega)$ we act
by the operator
$\left(\vec{p}\frac{\partial}{\partial\vec{k}}\right)^{2}$ on the
second Green's function in (2). The result reads
$$B=-\frac{\nu}{48\pi^{2}T^{2}}\sum_{n\geq0}\frac{1}{\left(n+\frac{1}{2}+\frac{\omega}{4\pi T}\right)^3}, \eqno (7) $$

The logarithmic divergence of the sum in (6) may be removed by the
introduction of the critical temperature $T_{c}$. First we
regularize (6) by introducing the high-frequency cut-off
$n_{max}=\frac{\Lambda}{2\pi T}$, $\Lambda\gg\omega$. Then

$$A=\nu\left\{\psi\left(\frac{1}{2}+\frac{\omega}{4\pi T}+\frac{\Lambda}{2\pi T}\right)-\psi\left(\frac{1}{2}+\frac{\omega}{2\pi T}\right)\right\}, \eqno (8)$$
where $\psi(z)$ is the logarithmic derivative of the
$\Gamma$-function. Next we replace $\Lambda$ by the critical
temperature $T_{c}$ using the relation

$$t=\ln\frac{T}{T_{c}}=\frac{1}{\nu g}-\psi\left(\frac{1}{2}+\frac{\Lambda}{2\pi T}\right)-\psi(\frac{1}{2}). \eqno (9)$$
Now we can return to the underlying formula (1) for the FQP, collect
all the pieces together and write

$$L^{-1}(\vec{p},\omega)=-\nu\left\{t+\psi\left(\frac{1}{2}+\frac{\omega}{4\pi T}\right)-\psi\left(\frac{1}{2}\right)-\frac{\vec{p}^{2}}{96\pi^{2}T^{2}}\psi''\left(\frac{1}{2}+\frac{\omega}{4\pi T}\right)\right\}, \eqno (10)$$
Equation (10) is the basic one in condenced matter fluctuation
theory [2]. We have shown that it can be almost literally retrieved
within rather general approach to dense finite temperature quark
matter. Fluctuations in quark matter are many orders of magnitude
stronger than in ordinary and even in high temperature
superconductors [1]. Quark matter formed in heavy ion collisions has
a finite volume which also increases fluctuation effects. Close to
the phase transition quark matter is a system with strong disorder
[1] possibly revealing Anderson localization. The FQP is known to be
an effective tool to study the properties of such systems. In
particular the poles of $L(\vec{p},\omega)$ determine the dynamical
diffusion coefficient. The detailed investigation of this problem is
beyond the scope of the present quick paper.

\section*{\normalsize{Acknowledgments}}

It is a pleasure to thank Tran Thanh Van for the invitation and warm
hospitality and Patricia Chemali and Bolek Pietrzyk for making my
participation possible. The support from RFBR qrants 07-02-08082,
06-02-17012, NSh-843.2006.2 is gratefully acknowledged.

\section*{\normalsize{References}}

1. B. Kerbikov and E. Luschevskaya, hep-ph/0607304, to be published
in Phys. Atom. Nucl., 2007.\\ 2. A. Larkin and A. Varlamov,
cond-mat/0109177, in: The Physics of Conventional and Unconventional
Superconductors. Eds. K. Bennermann,\\ J. Ketterson,
Springer-Verlag, Berlin-Heidelberg, 2001.

\end{document}